\documentclass[aps,pre,floats,twocolumn,showpacs,superscriptaddress]{revtex4}

\usepackage{natbib}
\usepackage[utf8]{inputenc}

\usepackage{amsmath, amsthm, amssymb}
\usepackage{enumitem}
\usepackage{graphicx}% Include figure files
\usepackage{times}% Times fonts
\usepackage{dcolumn}% Align table columns on decimal point
\usepackage{color}

\setenumerate{noitemsep,topsep=0pt,label=(\roman*)}

\begin{document}

\title{%
Absence of influential spreaders in rumor dynamics%
}

\author{Javier Borge-Holthoefer}
\email{borge.holthoefer@gmail.com}
\affiliation{Instituto de Biocomputaci\'on y F\'\i sica de Sistemas 
Complejos (BIFI), Universidad de Zaragoza, Mariano Esquillor s/n, 50018 Zaragoza, Spain}

\author{Yamir Moreno}
\email{yamir.moreno@gmail.com}
\affiliation{Instituto de Biocomputaci\'on y F\'\i sica de Sistemas 
Complejos (BIFI), Universidad de Zaragoza, Mariano Esquillor s/n, 50018 Zaragoza, Spain}
\affiliation{Departamento de F\'{\i}sica Te\'orica, Universidad de Zaragoza, 50009 Zaragoza, Spain}

\date{\today}

\begin{abstract}
Recent research \cite{kitsak2010identification} has suggested that coreness, and not degree, constitutes a better topological descriptor to identifying influential spreaders in complex networks. This hypothesis has been verified in the context of disease spreading. Here, we instead focus on rumor spreading models, which are more suited for social contagion and information propagation. To this end, we perform extensive computer simulations on top of several real-world networks and find opposite results. Namely, we show that the spreading capabilities of the nodes do not depend on their $k$-core index, which instead determines whether or not a given node prevents the diffusion of a rumor to a system-wide scale. Our findings are relevant both for sociological studies of contagious dynamics and for the design of efficient commercial viral processes.
\end{abstract}


\maketitle

%%%%%%%%%%%%%%%%%%%%%%%%%%%%%%%%
\section{Introduction}
Locating influential nodes in spreading dynamics is of utmost importance to understand and control these processes. In the case of epidemic spreading \cite{hethcote2000mathematics,pastor2001prl,gomez10epl}, the identification of the nodes responsible of system-wide disease outbreaks \cite{kitsak2010identification} has obvious advantages. For instance, this might allow minimizing the impact of the disease by dampening or removing the spreading capacity of influential nodes. Although rumor spreading processes are in some aspects similar to the propagation of epidemic diseases \cite{castellano2009statistical}, they are essentially designed to achieve the opposite goal, namely, to spread a piece of information (or news, opinions, etc) to as many nodes as possible. As a matter of fact, rumor-like mechanisms form the basis for many phenomena such as viral marketing, where companies exploit social networks to promote their products. The goal, then, is not to short-circuit the spreading process, but rather to explore ways to guarantee the highest outcome \cite{moreno2004dynamics,nekovee2007theory}. Thus, singling out ``privileged spreaders'' may be a way to maximize the chances to succeed in a marketing campaign at lower costs. 

Beyond commercial issues, recent social events also place the question of influential spreaders in the spotlight. All along 2011, we have witnessed an unusual amount of social unrest throughout the world. In the aftermath of financial and political crisis, and changes to welfare policies, protests have aroused in the form of pacific civil movements --the Spanish ``indignados'' in May or Occupy Wall-Street in the United States, culminated in global marches in October 15th.--, economy-related demonstrations with some violent episodes --Greece--, political (and sometimes violent) uprises  --the ``Arab spring''-- and still unclear riots in the United Kingdom in the last summer. In all of them, different online social sites (Facebook, Twitter, etc.) played an unprecedented key role at triggering collective phenomena and were used to help protesters disseminate their opinions and mottos to attain a critical mass of participants \cite{borge2011structural, gonzalez2011dynamics,borge2011locating} These new forms of social mobilization and protest demand new quantitative approaches to answer old sociological questions, such as how protest adherents share information and synchronize their activity at a global scale. 

In this paper, we focus on a sort of dynamics that can be used to answer this kind of questions. Specifically, we carry out extensive numerical simulations of a rumor spreading dynamics \cite{boccaletti06,moreno04efficiency} on a variety of real-world communication networks of sizes ranging from roughly $10^{3}$ to almost $10^{5}$ nodes. In doing so, we do not seek who is perceived by the network as an authority \cite{ratkiewicz2010characterizing, huang2011identifying}, or who is at the top of a hierarchical structure \cite{gupte2011finding}, but rather address the issue of influential nodes in a dynamical sense, i.e., we aim at identifying the nodes that play an outstanding role in the dissemination of information. This dynamical perspective is the one adopted in \cite{kitsak2010identification}, where the topological characteristics of nodes with special spreading capacities under epidemic dynamics (susceptible--infected--susceptible and susceptible--infected--removed models) were studied. The authors' findings indicate that centrality (measured through the $k$-core index \cite{alvarez2006large,alvarez2008k,carmi2007model}), and not degree, is the key topological feature to understand such spreading capacities. Capitalizing on this idea, we wonder whether there is a way to identify ``privileged spreaders'' also in rumor dynamics. Contrary to the results in \cite{kitsak2010identification}, we find that there are no such influential spreaders in rumor dynamics, rather there are nodes with an outstanding capacity to short-circuit the spreading process (henceforth referred to as {\em firewall} nodes). Besides, as far as the awareness of the rumor concerns, we show that the latter quantity is positively correlated with nodes' coreness.

The paper is organized as follows. In Sec.~\ref{sec:data} we first summarize the main characteristics of the networks on top of which the rumor dynamics is run. Next in Sec.~\ref{sec:model}, we detail which model of rumor spreading we shall be using to obtain the results presented in Sec.~\ref{sec:res}. Attending the previous distinction, we address the spreading and the forwarding/short-circuiting capacities separately. Our final conclusions are developed in Sec.~\ref{sec:conclusions}.
 
%%%%%%%%%%%%%%%%%%%%%%%%%%%%%%%%
\section{Datasets.}
\label{sec:data}

We have performed extensive numerical simulations of the dynamics of rumor spreading (presented in the next section) on top of several real-world networks. These networks are all relevant for the sort of dynamics we are dealing with. The networks used are:

\paragraph{Email contact network}
The network of email contacts is based on email messages sent and received within the Universitat Rovira i Virgili \cite{guimera03}. The data were collected in the time window between January and March of 2002. Nodes in the network represent email accounts, and an undirected link is placed whenever two email accounts exchanged emails, i.e. the lack of link direction reflects the fact that data is restricted to reciprocal communications. We consider only the giant connected component, which contains $N=1133$ nodes (out of a total of 1700) with an average degree of $\langle k \rangle = 9.6$.

\paragraph{Political blogs network}
Adamic and Glance \cite{adamic2005political} gave empirical evidence of the emergence of modular structure in the blogosphere around the specific topic of U.S. warfare policy, and the interaction of these communication tools with mainstream media. For this work they collected a directed network of hyperlinks between weblogs on U.S. politics, i.e. links stand for each time bloggers referred to one another. The giant connected component contains $N=1222$ with an average degree of $\langle k \rangle = 27.3$.

\paragraph{The Internet at the Autonomous Systems level (AS) network}
Nodes in the AS network are autonomous systems which are connected if there exists a physical connection between them. An AS is a connected group of one or more IP prefixes run by one or more network operators which has a single routing policy. Data have been recorded by CAIDA and corresponds to the snapshot of this structure of December 2009. The largest connected component of the AS network consists of 33235 ASs with an average degree $\langle k \rangle = 4.5$.

\paragraph{Twitter network}
Twitter is a microblogging site which, among other features, allows users to establish almost-static relations with other users by ``following'' them. This implies that if user $i$ follows $j$, $i$ will be informed of every message $j$ emits; in Twitter's jargon, $i$ is one of $j$'s follower. To extract this information for a subset of Twitter users in Spain \cite{gonzalez2011dynamics}, data were scrapped directly from {\em www.twitter.com} using a cloud of 128 different nodes of a subnet. The scrap was successful for 87569 users, for whom we obtained their official list of followers. It is worth remarking that the extraction of followers gave a list in the order of 3 millions users, which roughly coincides with the order of the audience estimated by Twitter in Spain; the list was however restricted to users who had some participation in the Spanish May 15th protests. The resulting structure is a directed network, direction indicates who follows who in the online social platform. Incoming links to a node $i$ signal which users $i$ is listening to, whereas out-going links point at those who are paying attention to (following) $i$.

We have also characterized the above networks according to their $k$-core index. This is done using the standard procedure \cite{alvarez2008k,carmi2007model}: particular subsets of the network, called $k$-cores, are each obtained by recursively removing all the vertices of degree less than $k$ (where $k = k_{in} + k_{out}$) until all vertices in the remaining graph have degree at least $k$. The higher the coreness of a node, the closer it is to the nucleus or core of the network.

%%%%%%%%%%%%%%%%%%%%%%%%%%%%%%%%
\section{Rumor dynamics}
\label{sec:model}

The model we shall consider is defined in the following way \cite{moreno04efficiency}. Each of the $N$ elements of the network can be in three possible states. We call a node holding an update and willing to transmit it a {\em spreader}. Nodes that are unaware of the update will be called {\em ignorants} while those that already know it but are not willing to spread the update anymore are called {\em stiflers}. We denote the density of ignorants, spreaders, and stiflers at time $t$ as $\psi(t)$, $\phi(t)$ and $s(t)$ respectively, such that $\psi(t) + \phi(t) + s(t) = 1$, $\forall t$. The spreading process takes place along the links between spreaders and ignorants. Each time step, spreaders contact one (or more) neighboring node. When the spreader contacts an ignorant, the last one turns into a new spreader at a rate $\lambda$. On the other hand, the spreader becomes a stifler with rate $\alpha$ if a contact with another spreader or a stifler takes place. This dynamics mimics the attempt to diffuse an update or rumor by nodes which have been recently updated. At the same time, if a node attempts too many times to communicate the update to nodes which have already received it, it stops the process, turning itself into a stifler. The dynamics terminates when the number of spreaders is 0, i.e. nobody in the network can circulate the rumor further.

From this general framework we have devised two versions of the dynamics. In the first one, at each time step the $\phi N$ spreaders are allowed to contact a single neighbor chosen at random (``contact process'', CP from now on). In the second, at each time step, each of the $\phi N$ spreaders contacts all its neighbors in a random sequence, unless during a contact it turns into a stifler. In this case it immediately stops contacting further nodes (``truncated process'', TP from now on). This accounts for the larger transmission capabilities of high degree nodes that can reach a larger number of neighbors as specified by the heterogeneous network topology. In both versions, the initial conditions are set such that  $\psi(0) = 1-1/N$, $\phi(0) = 1/N$ and $s(0)=0$. In addition, without loss of generality and unless explicitly stated, we fix from now on $\lambda = 1$ and $\alpha = 1$.

%%%%%%%%%%%%%%%%%%%%%%%%%%%%%%%%
\section{Results}
\label{sec:res}

Extensive numerical calculations were carried out simulating the dynamics of rumor propagation for both settings on top of the real-world networks previously introduced. From an initial scenario, in which all nodes belong to the ignorants class except the seed, we perform $S$ simulations. This is repeated for each node, i.e. every vertex of a network of $N$ nodes acts as the initial seed $S$ times, to obtain statistically significant results. $S$ is set to $10^{3}$ for the smallest networks, and to $10^{2}$ for the larger ones due to computational costs. In this way, for each node $i$, we average the final density of stiflers in the network $s^{i}_{\infty}$. This quantity accounts for the spreading capacity of node $i$, which quantifies how deep the rumor penetrated the network when node $i$ was the initial seed:

\begin{equation}
s^{i}_{\infty} = \frac{1}{S}\sum_{m=1}^{S}s^{i,m}_{\infty}
\end{equation}
where $s^{i,m}_{\infty}$ represents the final density of stiflers for a particular run $m$ with origin at node $i$. With this information at hand for all nodes, we coarse-grain the individual $s^{i}_{\infty}$'s into classes of nodes according to their degree. Thus, $M_{k_{S}}$ represents the average stifler density for all runs with a seed with a $k_{S}$ core index:

\begin{equation}
M_{k_{S}} = \sum_{i \in \Upsilon_{k_{S}}} \frac{s^{i}_{\infty}}{N_{k_{S}}}
\end{equation}
where $\Upsilon_{k_{S}}$ is the set of all $N_{k_{S}}$ nodes with $k_{S}$ values. Similarly, the final density of stiflers obtained when node $i$ acts as the initial spreader is averaged over all the seeds with the same $(k, k_{S})$ values \cite{kitsak2010identification}:

\begin{equation}
M_{k, k_{S}} = \sum_{i \in \Upsilon_{k_{S}, k}} \frac{s^{i}_{\infty}}{N_{k_{S}, k}}
\end{equation}
where $\Upsilon_{k_{S}, k}$ is the set of all $N_{k_{S}, k}$ nodes with $(k, k_{S})$ values.

\subsection{Spreading capability}

\begin{figure}
  \centering
  \includegraphics[width=\columnwidth,clip]{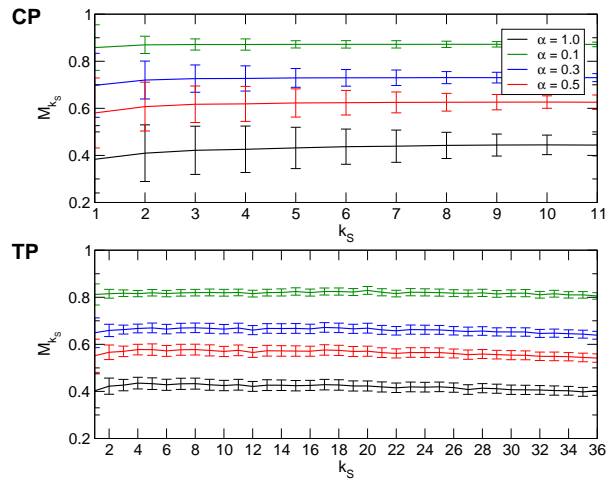}
  \caption{(color online) Average stifler density for rumor processes initiated at nodes with coreness $k_S$. The top panel shows the results obtained for the CP run on top of the email network, while the bottom panel represents the same dependency but for the TP on top of the political blogs network. As it can be seen from the figures, no matter the value of $\alpha$, there is no correlation between $M_{k_{S}}$ and the seeds centrality, which leads to the absence of influential spreaders in rumor dynamics.}
  \label{fig1}
\end{figure}

\begin{figure*}
  \centering
  \includegraphics[width=\textwidth,clip]{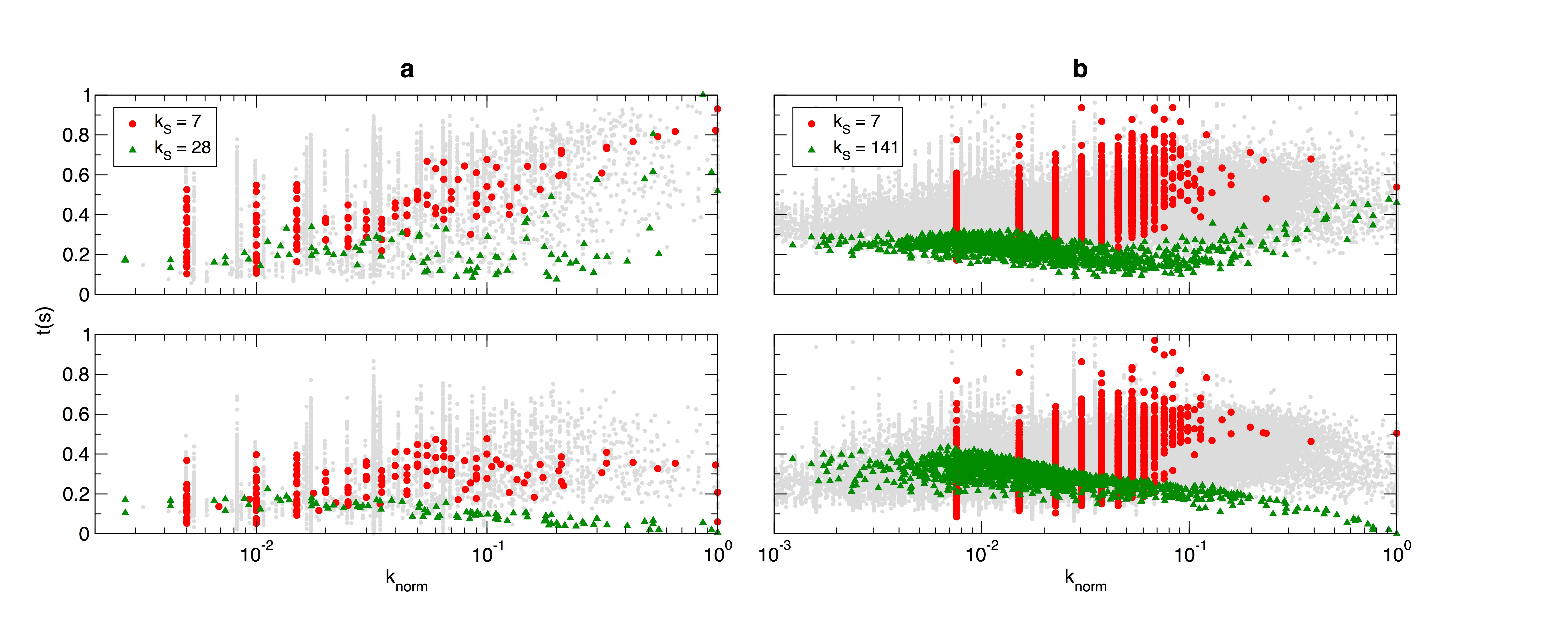}
  \caption{(color online) Central nodes (those in the highest $k$-shell) act as firewalls of the rumor spreading. This hypothesis is validated by the fact that they are among the first to become stiflers (low values of $t(s)$), thus acting as topological barriers for the dynamics. Within these central nodes, the time at which they turn into stiflers is even earlier for those with highest degree --the exception to this being a small fraction of them in the CP setting (upper panels). See main text for details.}
  \label{fig2}
\end{figure*}

Figure ~\ref{fig1} shows results from simulations of the rumor dynamics using both the contact and the truncated settings on top of two of the networks used (similar results are obtained for the rest of networks). As it can be seen, $M_{k_{S}}$ is almost invariant regardless of the core number of the seed node. In other words, the spreading capabilities of the different nodes are almost the same wherever the dissemination begins. This implies that the topological position of nodes has no impact on the final success of a rumor, revealing an absence of influential spreaders for rumor dynamics. 

This counterintuitive result is in sharp contradiction with results obtained for epidemic dynamics in \cite{kitsak2010identification}, and underlines the fact that, beyond shallow similarities with such processes (specially in reference to the SIR scheme), rumor dynamics display a markedly different behavior. In an attempt to understand these differences, we hypothesize that barriers to information dissemination must be placed in topologically important nodes. In other words, the dynamics favors that central nodes (as given by the connectivity) turn into the stifler class \cite{moreno04efficiency}, thus short-circuiting any further expansion of the rumor. Be this true, such firewall nodes must belong to the subset of the population which become stiflers more quickly. This amounts to say that nodes belonging to a higher core should promptly reach the stifler class. Furthermore, $k$-shells are built up of nodes with heterogeneous degrees, thus nodes with higher degrees should also turn to stiflers more rapidly than lower degree nodes within the same class $k_{S}$. Figure~\ref{fig2} represents this situation for the largest networks analyzed in this work (AS, left panels {\bf a}; Twitter, right panels {\bf b}). This plot shows, for each node in the network, the average time it takes to become a stifler $t(s)$. To grasp in a single figure the influence of the core level $k_{S}$ and degree heterogeneity within these levels, we have normalized $k$. That is, for a node belonging to the shell $S$, its normalized degree is measured as 
\begin{equation}
k_{norm} = \frac{k_{i} - k_{min}^{S}}{k_{max}^{S} - k_{min}^{S}}
\end{equation}
where $k_{min}^{S}$ is the minimum degree in shell $S$, and $k_{max}^{S}$ is the maximum. In doing so, we can overlap all $t(s)$ on a common interval $[0, 1]$, so we can easily check if the hypotheses actually hold. The time $t(s)$ has also been normalized in Figure~\ref{fig2} to allow comparison across dynamics (CP and TP, upper and lower panels respectively) and between networks. For the sake of clarity, we have highlighted only two $k$-shells, a peripheral one (i.e., low $k_{S}$, red circles) and that with the highest index ($k_{S_{max}}$, green triangles).

Figure~\ref{fig2} reveals that indeed nodes in the highest core turn to stiflers quite early, compared to the rest of the network. Also, the main trend within the highest core is that $t(s)$ decreases as the degree of nodes belonging to that core increases. This is clearly seen specially for the TP setting (lower panels), but also in the CP dynamics except for some nodes with highest degrees (right-most dots). These statements are even more clear when confronted to results corresponding to nodes in the 7th shell (red circles), which shows (with few exceptions) systematically larger $t(s)$s. Therefore, the absence of influential spreaders in rumor dynamics is translated into the existence of critical firewalls that do depend on the coreness of the nodes.

\subsection{Awareness of information}

\begin{figure*}
  \centering
  \includegraphics[width=\textwidth,clip]{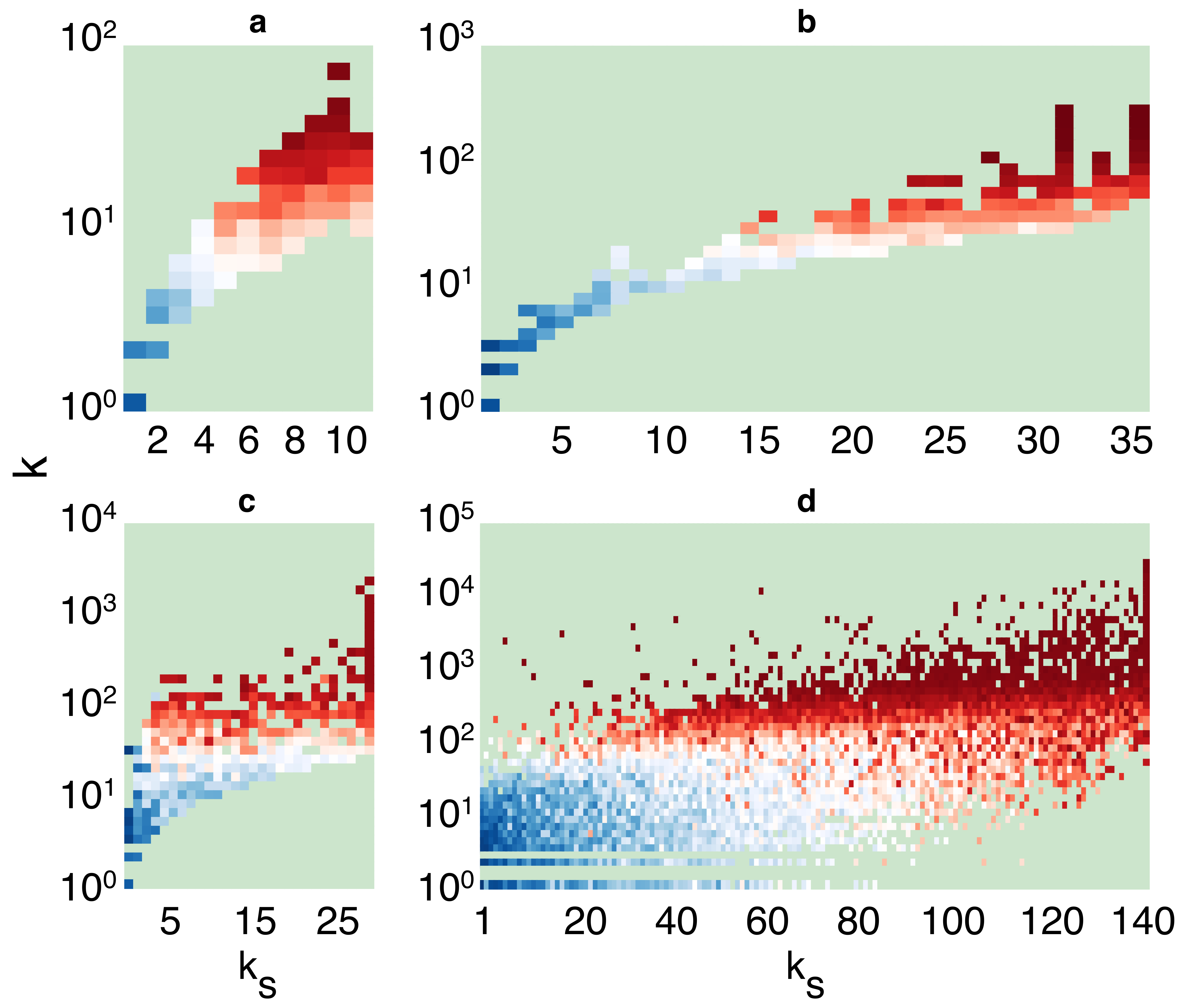}
  \caption{(color online) Awareness level, i.e., percentage of nodes that have heard the rumor, for each class $(k, k_{S})$ for the four networks used in this work $-$a) email network, b) political blogs network, c) AS network and d) Twitter network. We have represented the results obtained for the TP, but the CP dynamics give qualitatively similar results. Nodes with low degree (compared to $k_{max}$) but mid-to-high coreness are systematically reached by roughly any generated rumor, regardless of its origin (colors white to red correspond to an awareness level above 50\%, while light green indicates that no nodes belong to that class. See the text for further details.}
  \label{fig3}
\end{figure*}

The previous results suggest a different approach to the concept of influential nodes in rumor spreading dynamics. The existence of critical firewalls demands their turning to stiflers, which in turn implies their being spreaders in the first place. If critical firewalls can systematically short-circuit the dissemination of information, it is only because they can actually ``hear'' any rumor circulating in the network. Thus, beyond the capacity of a seed to generate system-wide information cascades, one must also take into account who in the network keeps track of all the information. For instance, in the context of social protest and unrest, agents who have access to most of the information may have an important role in filtering knowledge (perhaps acting as information sinks) and in coordinating and synchronizing the collective action. Actually, empirical data evidences that these critical nodes may act as enhancers sometimes, and as firewalls others \cite{borge2011structural}.

Figure~\ref{fig3} illustrates the previous idea. The different panels represent the proportion of nodes, $M_{k, k_{S}}$, that have heard the rumor at the end of the process. The results depicted in all panels suggest that, in most cases, centrality (high core) is a sufficient condition to become aware of the rumor in at least half the number of processes simulated (cold colors indicate low awareness, less than 50\%, whereas colors in the scale from white to red indicate a fraction larger than 50\% of heard messages). Unsurprisingly, nodes with large $k$ are also aware of most rumors because of the many paths leading to them. In this sense, we find that the coreness is a better descriptor than the degree when it comes to quantify how likely is that a given node be in the stifler class at the end of the rumor dynamics.

%%%%%%%%%%%%%%%%%%%%%%%%%%%%%%%%
\section{Conclusions}
\label{sec:conclusions}
Recent events throughout the world involving civil movements and collective action --critically aided by online social networks-- have called attention to the problem of identifying influential spreaders. This question has many facets (hierarchies, authorities, prestige measures), but Kitsak {\em et al.} \cite{kitsak2010identification} have addressed it linking a vertex topological feature ($k$-core) with its dynamic behavior in a simple and operative way. However, previous works have only dealt with epidemic dynamics, which do not grasp all phenomena at stake when it comes to information dissemination. In this work we present an alternative dynamical process which is more suited to describe contagious dynamics that are mediated by information spreading. Taking four real-world communication networks in which such dynamical processes may take place, we have performed extensive numerical simulations of two versions of a rumor dynamics \cite{moreno04efficiency,moreno2004dynamics,nekovee2007theory}.

Three main conclusions stem from this work. First, the rumor dynamics does not favor the appearance of influential spreaders. No matter where a rumor is triggered, the fraction of stifler nodes is similar for the same values of the parameters. Second, the absence of such privileged spreaders does not imply that central nodes (may them belong to a high $k$-shell or have a large $k$) are not important in the dynamics. However, such distinct role is not as {\em enhancers} of the dissemination, rather they behave like firewalls that interrupt the propagation of information. Indeed, it follows that the $k$-core index is a useful predictor of different roles in the context of rumor dynamics, but it is paradoxically in the opposite direction of what one would expect. The short time needed by topologically outstanding nodes to turn into stiflers is a good indicator of how they choke the spreading at early times --and they do so regardless of the seed, hence the observed uniform spreading capacity. This result agrees with empirical facts recently reported: hubs may play the role of information filters or sinks \cite{borge2011structural}. 

Finally, in connection to the previous point, the fact that central nodes act as rumor firewalls implies that many rumors reach them before dying out. This means that, although there are no privileged spreaders, nodes in high core levels have information advantage in the form of awareness, i.e. they have access to all the information circulating around the network. We believe that research on rumor models can be taken further. Data from online social networks offers unprecedented opportunities to gain insight in how --and where-- information is generated, exchanged, forwarded or banned. From this evidence we might devise more realistic dynamics that increase our understanding of this phenomenology.

%%%%%%%%%%%%%%%%%%%%%%%%%%%%%%%%
\section*{Acknowledgments}
This work has been partially supported by MICINN through Grants FIS2008-01240 and FIS2009-13364-C02-01, and by Comunidad de Arag\'on (Spain) through a grant to FENOL group.

%\bibliography{bibtesi}

\end{document}